\documentclass[]{raa}
\usepackage{graphicx,times}
\usepackage{natbib}

\def\LIR{L_{\rm IR}}
\def\Lsun{L_\odot}

\begin{document}

\title{The Evolution of Advanced Merger (U)LIRGs on the Color-Stellar Mass Diagram}

\volnopage{Vol.0 (200x) No.0, 000--000}      
\setcounter{page}{1}          

\author{Rui Guo
   \inst{1, 2, 3}
 \and Cai-Na Hao
   \inst{3}
 \and Xiao-Yang Xia
   \inst{3}}
 \institute{National Astronomical Observatories, Chinese Academy of Sciences,
            20A Datun Road, Chaoyang District, Beijing 100012, China\\
      \and
            University of Chinese Academy of Sciences, Beijing 100049, China\\
      \and
            Tianjin Astrophysics Center, Tianjin Normal University, Tianjin
            300387, China; {\it E-mail: cainahao@gmail.com}\\
           }
\date{Received~~ ; accepted~~ }

\abstract{
Based on a sample of 79 local advanced merger (adv-merger) (U)LIRGs, we search 
for the evidence of quenching process by investigating the
distributions of the star formation history indicators (EW(H$\alpha$),
EW(H$\delta$$_A$) and D$_n(4000)$) on the {\em NUV-r} color-mass and
SFR-$M_{\ast}$ diagrams.  The distributions of the EW(H$\alpha$) and
D$_n(4000)$ on the {\em NUV-r} color-mass diagram show clear trends that at a
given stellar mass, galaxies with redder {\em NUV-r} colors have smaller
EW(H$\alpha$) and larger D$_n(4000)$. The reddest adv-merger (U)LIRGs close to
the green valley have D$_n(4000)$$>1.4$ mostly. In addition, in the SFR-$M_{\ast}$ diagram,
as the SFR decreases, the EW(H$\alpha$) decreases and the D$_n(4000)$ increases, implying
that the adv-merger (U)LIRGs on the star formation main sequence have more evolved stellar
populations than those above the main sequence. These results indicate that a fraction
of the adv-merger (U)LIRGs have already exhibited signs of fading from the
starburst phase and that the {\em NUV-r} reddest adv-merger (U)LIRGs are likely at
the initial stage of post-starbursts with age of $\sim 1$ Gyr, which is
consistent with the gas exhausting time-scales. Therefore, our results offer
additional support for the fast evolutionary track from the blue cloud to the red sequence.
}

   \authorrunning{Guo et al.}
   \titlerunning{Evolution of Advanced Merger (U)LIRGs}

   \maketitle

\section{INTRODUCTION}

Since the discovery of the color bimodality on the color-magnitude and the
color-stellar mass (color-mass, hereafter) diagrams based on the Sloan Digital
Sky Survey (SDSS) (Kauffmann et al. 2003b; Baldry et al.  2004, 2006) as well
as on high redshift survey samples (Bell et al. 2004; Faber et al.  2007;
Ilbert et al. 2010), there are mounting works to investigate the evolutionary
pathways from the blue cloud of star-forming galaxies to the red sequence of
quiescent galaxies.  The evolution from the blue cloud to the red sequence is
suggested to be results of star formation being quenched (e.g., Faber et al.
2007). According to the quenching time-scales, the evolutionary tracks are
classified into fast and slow modes (e.g., Schawinski et al. 2014; Yesuf et al. 2014).  
The fast track (with time-scales less than 1\,Gyr) has been proposed to interpret the
formation of high redshift compact massive quiescent galaxies (van Dokkum et al. 2008;
Kriek et al.  2009; Barro et al. 2013; Muzzin et al. 2013; Marchesini et al. 2014).
These studies show that the progenitors of local massive galaxies are heavily
dust-extincted starburst galaxies that are triggered by gas-rich major mergers.
Therefore, investigation of local gas-rich major mergers might shed
light on the fast quenching process.

In the local universe, almost all ultraluminous infrared galaxies (ULIRGs;
$\LIR$\footnote{$\LIR$ is the integrated infrared luminosity between 8-1000
$\mu$m.}$>10^{12}\Lsun$) and about half of luminous infrared galaxies (LIRGs;
$10^{11}\Lsun<\LIR<10^{12}\Lsun$) were found to be gas-rich interacting/merging
galaxies (e.g. Sanders \& Mirabel 1996; Wang et al. 2006; Kaviraj 2009).
Therefore, nearby (U)LIRGs can serve as a proper local laboratory for studying
the fast quenching process.  Virtually, several recent works have made efforts
to identify the role of local ULIRGs in the migration from the blue cloud to
the red sequence. Chen et al. (2010) and Kilerci Eser et al.  (2014) have
investigated the positions of local ULIRGs in the {\em g-r} and {\em u-r}
color-magnitude diagrams, respectively, without applying internal extinction
corrections. They found that about half of the ULIRGs sample lie outside
the 90\% level number density contour, and some in the green valley or the red
sequence. Most recently, Guo et al. (2016b) applied both Galactic and internal
dust extinction corrections for an advanced merger (adv-merger) (U)LIRGs sample and found that
adv-merger (U)LIRGs are blue and more massive than the blue cloud galaxies,
with $95\%\pm 2\%$ and $81\%\pm 4\%$ of them outside the blue cloud on the {\em
u-r} and {\em NUV-r} color-mass diagrams, respectively, implying that the
adv-merger (U)LIRGs occupy a distinct region on the color-mass diagrams. In
addition, the investigation of the locus of the adv-merger (U)LIRGs in the star
formation rate (SFR)-stellar mass (SFR-$M_{\ast}$) diagram showed that about
three-fourths of the sample are above the star-forming main sequence, indicating
that the majority of adv-merger (U)LIRGs are experiencing massive starbursts.
Based on the estimates of gas exhausting time-scales, the authors suggested
that the adv-merger (U)LIRGs are likely at the starting point of the fast
evolutionary track.

As is well known, the equivalent width of H$\delta$$_A$ absorption line
(EW(H$\delta$$_A$)) and the 4000{\rm \AA} break (D$_n(4000)$) measured from
optical spectra are powerful indicators of the recent star formation histories.
A strong H$\delta$$_A$ absorption line is a sign for a burst in the star
formation history that ended about 0.1-1 Gyr ago, while the 4000{\rm \AA} break
is an excellent age indicator for young star populations within 1 Gyr (Kauffmann
et al. 2003b, 2003c). As a result, EW(H$\delta$$_A$) and D$_n(4000)$ are often used to
probe recent starbursts and distinguish star formation histories
dominated by starbursts from a continuous mode (Kauffmann et al. 2003b; Goto
2007; Martin et al. 2007; Yesuf et al. 2014).  In addition, the equivalent
width of H$\alpha$, EW(H$\alpha$), is a measure of the current to past average SFR
(e.g., Knapen \& James 2009; Yesuf et al. 2014).

To explore the evolutionary link between starburst galaxies and
post-starbursts, Yesuf et al. (2014) proposed a classification scheme to
identify transiting post-starbursts, an intermediate stage between starbursts
and post-starbursts, based on EW(H$\alpha$), EW(H$\delta$$_A$), D$_n(4000)$, UV
and {\em Wide-field Infrared Survey Explore} ({\em WISE};
Wright et al. 2010) photometry. Guo et al. (2016b) employed this set of criteria
and identified 12 transiting post-starbursts out of the entire sample of 89
adv-merger (U)LIRGs. It is also worth carefully investigating
the distributions of the spectral features of EW(H$\alpha$),
EW(H$\delta$$_A$), D$_n(4000)$ on the color-mass and SFR-mass diagrams to
search for evidence for possible evolutionary signs already shown in the
adv-merger (U)LIRGs sample.

The paper is structured as follows. In Sections 2 we describe the
sample selection and parameter estimations for our sample adv-mergers and
control samples.  In Section 3, we present our results and discussion. The 
main results are summarized in Section 4. Throughout this paper we adopt the Kroupa
(2001) initial mass function and a cosmological model of ${H}_{\rm
0}=70\,{\rm km \, s^{-1}\, Mpc^{-1}}$, $\Omega_{\rm m}=0.3$ and $\Omega_{\rm
\Lambda}=0.7$.

\section{SAMPLE SELECTION AND PARAMETER ESTIMATIONS}

\subsection{Sample}

Our adv-merger (U)LIRGs sample is directly taken from Guo et al. (2016b). 
We refer the readers to that paper for a detailed description of the
sample selection (see also Guo et al. 2016a). In the following, we briefly describe the selection procedures.

\begin{figure}
\includegraphics[width=\textwidth, angle=0]{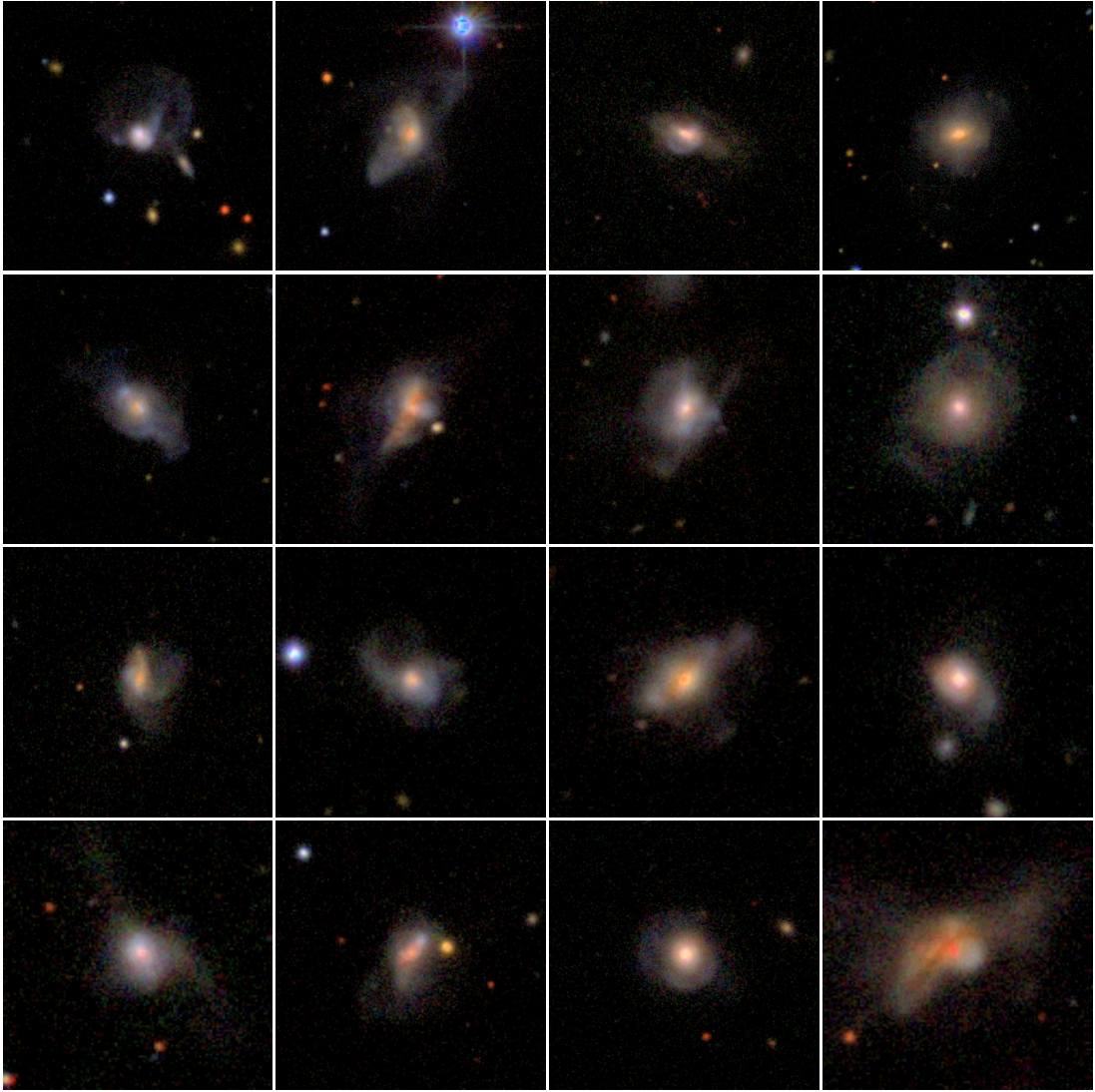}
\caption{RGB color images of 16 example adv-mergers with single nucleus but
some merger signatures, e.g., tidal tails, indicating that they are in the
late stage of merging. The images are constructed from SDSS
({\em g, r, i}) images following Lupton et al. (2004) with
80 $\times$ 80 $\rm kpc^2$ physical size.}
\label{colorim.eps}
\end{figure}

The parent (U)LIRGs sample in Guo et al. (2016b) was drawn from a
cross-correlation analysis between the spectroscopic catalog of SDSS DR7 (York
et al. 2000; Abazajian et al. 2009) and the {\em Infrared Astronomical
Satellite (IRAS)} Point Source Catalog Redshift Survey (PSCz, Saunders et al.
2000), as well as a cross-identification between the {\em IRAS} 1 Jy ULIRGs
sample (Kim \& Sanders 1998) and the SDSS DR7 photometric catalog.  We
performed morphological classifications visually for the sub-sample of galaxies
in the magnitude range of $14.5<r<15.9$\footnote{The magnitude restriction was
not performed for the 1 Jy ULIRGs since they have obvious interacting or
merging morphologies.} after corrections for foreground Galactic extinction
(Schlegel et al. 1998), to ensure both reliable morphological classifications
and the spectroscopic observation completeness for SDSS galaxies (Fukugita et
al. 2004; Kauffmann et al. 2003b).  Note that galaxies classified as Seyfert I
are removed from our sample, since the estimations for their extinctions,
stellar masses and star formation rates are not reliable.  Finally, there are
89 adv-mergers selected, which are in the late merging stage.  For the purpose
of this work, measurements of optical spectral indices are needed. Hence only
the 79 adv-mergers with SDSS spectroscopic observations are included in this
paper, among which 59 are LIRGs and 20 are ULIRGs.  Figure \ref{colorim.eps}
shows the true color images of 16 example adv-mergers. 

Meanwhile, the control samples are also taken from Guo et al. (2016b) directly.
For the study of the local color-mass relation, the control sample was selected
from the Oh-Sarzi-Schawinski-Yi (OSSY) catalog (Oh et al.  2011)\footnote{
http://gem.yonsei.ac.kr/$\sim$ksoh/wordpress}, which provides internal
extinction information (E(B-V)) from stellar continuum fits.  By requiring the
redshift range of $0.02<z<0.05$ and absolute magnitude $M_{z,\rm Petro} < -19.5$ mag,
53604 galaxies were obtained.  The absolute magnitude constraint enables us to derive
an approximately mass-limited sample as described in Schawinski et al. (2014).
For investigating the SFR-$M_{\ast}$ relation of star-forming galaxies, the
control sample has been retrieved from the SDSS DR7 in the redshift range of
$0.005<z<0.2$ and luminosity range of $14.5<r<17.77$, which consists of 152137
galaxies.

\subsection{Parameters}

The main purpose of this paper is to investigate the distributions of
EW(H$\alpha$), EW(H$\delta_A$) and D$_n(4000)$ on the {\em NUV-r} color-stellar
mass as well as the SFR-stellar mass diagrams. Therefore, apart from the estimates of the
dust extinction corrected {\em NUV-r} colors, SFRs and stellar
masses, as described in Guo et al. (2016b), optical spectral indices are also needed.

The estimations of the dust attenuation corrected {\em NUV-r} colors, SFRs and
stellar masses have been described in Guo et al. (2016b). We only provide a brief
description here.  The optical photometric data and {\em NUV} magnitudes were
taken from the SDSS DR7 and the {\em Galaxy Evolution Explorer (GALEX)} satellite (Martin et al. 2005),
respectively.  The {\em NUV} photometric data are available from the {\em
GALEX} for 69 sample adv-merger (U)LIRGs and $\sim87\%$ control sample galaxies.  We
further performed {\em k}-corrections, the Galactic extinction and internal
extinction corrections to the optical and {\em NUV} magnitudes.  The stellar
masses for our sample adv-mergers and the control sample galaxies were
retrieved from the Max Planck Institute for Astrophysics-Johns Hopkins
University (MPA/JHU\footnote{http://www.mpa-garching.mpg.de/SDSS}) stellar mass
catalog (Kauffmann et al. 2003b), which are based on the SDSS five broad-band
photometry. The SFRs for 25 star-forming and 38 composite\footnote{The spectral
classification were performed according to the BPT diagram (Baldwin et al.
1981; Kauffmann et al. 2003a; Kewley et al. 2001).} adv-merger (U)LIRGs were calculated
using their H$\alpha$ luminosities taken from the MPA/JHU catalog following
Kennicutt (1998) and then were converted to a Kroupa IMF. Note that the
H$\alpha$ luminosities are aperture-corrected (Hopkins et al. 2003) as well as
dust extinction corrected under the assumption of the case B recombination
value of intrinsic H$\alpha$/H$\beta$ as 2.86.  

The D$_n(4000)$, EW(H$\alpha$) and EW(H$\delta$$_A$) were retrieved from the
MPA/JHU catalog, in which the EW(H$\delta$$_A$) has been corrected for the
contamination by nebular emission.  We further performed the internal
extinction corrections for D$_n(4000)$ and EW(H$\alpha$), using the Calzetti's
law (Calzetti et al. 2000).

\section{RESULTS AND DISCUSSION}

As presented in Guo et al. (2016b), most of our sample adv-merger (U)LIRGs are
located above the star-forming main sequence on the SFR-$M_{\ast}$ diagram, 
indicating that they are experiencing massive starbursts as triggered by gas-rich major merging.  
All these galaxies are blue and massive, and thus lie to the right of the blue
cloud galaxies on the optical and NUV color-mass diagrams, making them distinct
from the blue cloud, red sequence and green valley galaxies. These properties
suggest that they are at the starting point of the fast track of galaxy
evolution from the blue cloud to the red sequence. Therefore, it is interesting to ask
whether they have already shown an evolutionary trend, as imprinted in their star
formation history indicators. In this work, we investigate the distributions
of the star formation history indicators, EW(H$\alpha$), EW(H$\delta$$_A$) and
D$_n(4000)$ on the color-mass and SFR-mass diagrams to search for evidence of 
possible evolution among the adv-merger (U)LIRGs sample.

\subsection{Distributions of EW(H$\alpha$), EW(H$\delta$$_A$) and D$_n(4000)$ in the {\em NUV-r} Color-Mass Diagram \label{subsec:colormass}}

\begin{figure}
\includegraphics[width=\textwidth, angle=0]{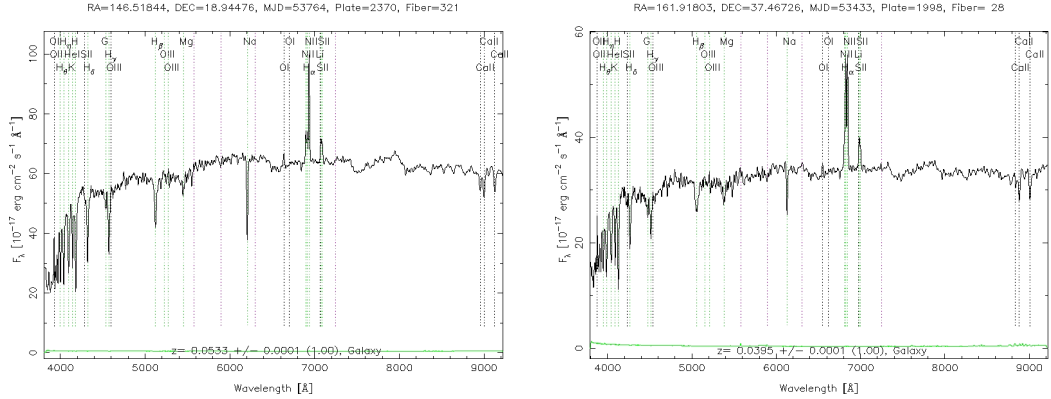}
\caption{SDSS spectra for 2 example adv-merger (U)LIRGs with weak H$\alpha$
emissions, strong H$\delta_{A}$ absorptions and deep 4000$\rm \AA$ breaks.}
\label{spectra.eps}
\end{figure}

Although a large fraction of our sample adv-merger (U)LIRGs are undergoing
massive starbursts, the optical spectra of several galaxies in our sample already
present clear K+A like features. Figure \ref{spectra.eps} shows the example spectra of two such
galaxies, from which we can clearly see the K+A like spectroscopic signatures, i.e.,
weak H$\alpha$ emission, strong H$\delta$$_A$ absorption line and deep 4000{\rm
\AA} break. These features indicate that a small fraction of adv-merger (U)LIRGs
are more evolved than the others under the assumption that the adv-merger (U)LIRGs
have similar star formation histories.

\begin{figure}
\includegraphics[width=\textwidth, angle=0]{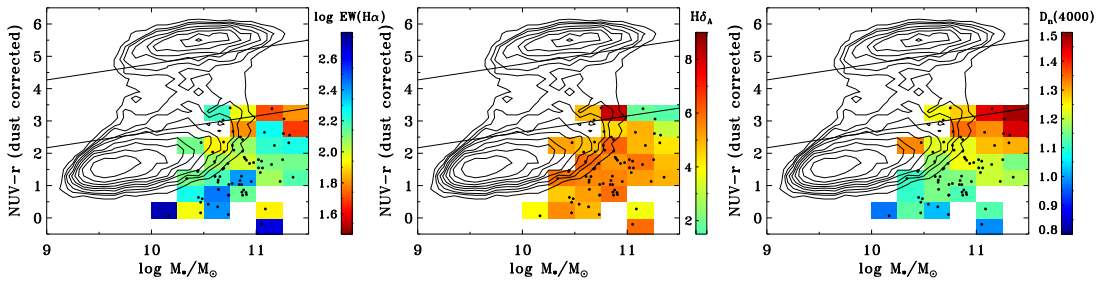}
\caption{Dust-corrected {\em NUV-r} color-mass diagrams for 69 sample
adv-merger (U)LIRGs with both {\em NUV} photometry from the {\em GALEX} and
spectra from the SDSS. The contours represent 9 equally spaced levels between
10\% and 90\% number densities of 46407 control sample galaxies with {\em NUV}
photometry.  Black solid lines show the boundaries of the green valley
(Equations (7) and (8) from Guo et al. (2016b)).  Dust extinction corrected
EW(H$\alpha$) (left), EW(H$\delta_{A}$) (middle) and D$_n(4000)$ (right)
distributions are overlaid. The color scales are shown to the right of the
respective panel.
}
\label{cmd.eps}
\end{figure}

To get a visual view of the evolution, we use different colors to represent
the EW(H$\alpha$), EW(H$\delta$$_A$) and D$_n(4000)$ values on the {\em NUV-r}
color-mass diagram, shown in the left, middle and right panels of Figure
\ref{cmd.eps}, respectively. It is clear from the middle panel of Figure
\ref{cmd.eps} that the EW(H$\delta$$_A$) values for most sample galaxies are
larger than 4{\rm \AA}, indicating A-type star features from a burst in the
recent past. In the left and right panels of Figure \ref{cmd.eps}, there are
obvious trends that the redder the galaxies are on the {\em NUV-r} color-mass
diagram, the smaller the EW(H$\alpha$) and the larger the D$_n(4000)$ are. The
reddest adv-merger (U)LIRGs close to the green valley mostly have
D$_n(4000)$$>1.4$.  We note that such trends are not caused by the correlation
between the stellar mass and the EW(H$\alpha$) or D$_n(4000)$, as manifested by
the apparent trends shown at a given stellar mass.  Taking into account the
evolutionary picture as traced by the D$_n(4000)$ and EW(H$\delta$$_A$) for an
instantaneous starburst by Kauffmann et al. (2003b), the adv-merger (U)LIRGs
with redder {\em NUV-r} colors are likely close to the end of starbursts or at
the initial stage of post-starbursts according to the definitions by Goto
(2007), Martin et al.  (2007) and Yesuf et al. (2014).  Therefore, from the
distributions of the EW(H$\alpha$), EW(H$\delta$$_A$) and D$_n(4000)$ values on
the {\em NUV-r} diagram in Figure \ref{cmd.eps}, we can clearly see the
evolutionary trend from starbursts with blue {\em NUV-r} colors ($\sim 0.5$
mag) to relatively older populations with redder colors ($\sim 3$ mag) that
formed in recent starbursts within the past $\sim 1$ Gyr, which is consistent
with the gas exhausting time-scales as estimated in Guo et al. (2016b). This
adds additional evidence for the fast evolutionary pathway to the conclusion
drawn by Guo et al. (2016b).

\subsection{Distributions of EW(H$\alpha$), EW(H$\delta$$_A$) and D$_n(4000)$ in the SFR-$M_{\ast}$ Diagram}

It has been accepted that the SFR-$M_*$ relation reflects the star formation
modes: galaxies on the star formation main sequence form stars in a relatively
steady mode, while galaxies localized above the main sequence line form stars in a
starburst mode (Rodighiero et al. 2011; Hung et al. 2013). From the analysis
for adv-merger (U)LIRGs, Guo et al. (2016b) found that about three-fourths of
adv-merger (U)LIRGs are located above the $1\, \sigma$ line of the local
star-forming galaxy main sequence and one-fourth of the sample galaxies lie on
the main sequence. Therefore, it is essential to investigate the reasons
responsible for the difference between the galaxies above the main sequence and
those on the main sequence. On the other hand, it is a complement to Section
\ref{subsec:colormass} to witness the evolution of the stellar populations 
on the SFR-$M_{\ast}$ diagram.
 
\begin{figure}
\includegraphics[width=\textwidth, angle=0]{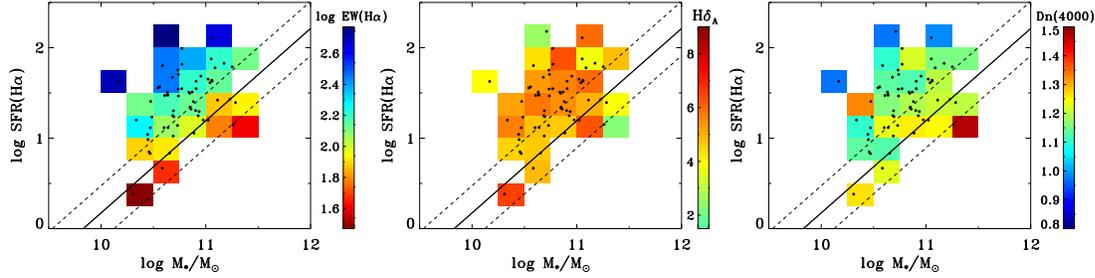}
\caption{SFR(H$\alpha$) vs. stellar mass relations for 63 star-forming and
composite sample adv-merger (U)LIRGs.
The black solid line indicates the local star-forming main sequence (MS) (Equation
(4) from Guo et al. (2016b)) with 1 $\sigma$ dispersions (0.3 dex) shown as
dashed lines.
Dust extinction corrected EW(H$\alpha$) (left), EW(H$\delta_{A}$) (middle) and
D$_n(4000)$ (right) distributions are overlaid. The color scales are shown to
the right of the respective panel.}
\label{ms.eps}
\end{figure}

Similar to Figure \ref{cmd.eps},  different colors are used to denote the
EW(H$\alpha$), EW(H$\delta$$_A$) and D$_n(4000)$ values for each sample galaxy
on the SFR-$M_{\ast}$ diagram, shown in the left, middle and right panels of Figure
\ref{ms.eps}, respectively. It is clear from the left and right panels of 
Figure \ref{ms.eps} that at a given stellar mass, there are clear trends 
that as the SFR decreases, the EW(H$\alpha$) decreases and the
D$_n(4000)$ increases. In addition, the sample galaxies on the main
sequence have relatively smaller EW(H$\delta$$_A$) on average. These results 
suggest that the adv-merger (U)LIRGs localized on the main sequence are more evolved
counterparts of those above the main sequence, and they have started to fade
from the starburst stage, as indicated by their smaller EW(H$\alpha$) and
larger D$_n(4000)$. According to the starburst age estimated by D$_n(4000)$
(Kauffmann et al. 2003b), these adv-merger (U)LIRGs are going to leave the main
sequence towards low SFRs in the SFR-$M_{\ast}$ diagram in $\sim 1$ Gyr. The
evolution of the spectral indices on the SFR-$M_{\ast}$ diagram depicts a coherent
picture with that exhibited by Figure \ref{cmd.eps}.
  
\section{SUMMARY}

With the aim of studying the fast quenching process from the blue cloud to the
red sequence, we retrieve 79
local adv-merger (U)LIRGs with SDSS spectroscopic observations from Guo et al.
(2016b) and investigate the distributions of EW(H$\alpha$), EW(H$\delta$$_A$)
and D$_n(4000)$ on the {\em NUV-r} color-mass and SFR-$M_{\ast}$ diagrams.  The
distributions of EW(H$\alpha$), EW(H$\delta$$_A$) and D$_n(4000)$ values on the
{\em NUV-r} color-mass and SFR-$M_{\ast}$ diagrams of 
adv-merger (U)LIRGs show obvious evolutionary trends:
at a given stellar mass, galaxies with redder {\em NUV-r} colors or
lower SFRs have smaller EW(H$\alpha$) and larger D$_n(4000)$, indicating more
evolved stellar populations. Moreover, most of the reddest adv-merger (U)LIRGs close to
the green valley have D$_n(4000)$$>1.4$ and the majority of our sample galaxies
have EW(H$\delta$$_A$) $> 4$\AA\ that is a sign for A-type star features from a recent
starburst in the past $0.1-1$ Gyr.  These results suggest that some of the
adv-merger (U)LIRGs have already shown signs of fading from the starburst stage 
and the reddest adv-merger (U)LIRGs in {\em NUV-r} color are likely at the
initial stage of post-starbursts with age of $\sim 1$ Gyr. This time-scale
coincides with the gas depletion time-scales derived by Guo et al. (2016b). 
Therefore, our results add additional support for the fast evolutionary track
proposed by several groups (e.g.  Muzzin et al.  2013; Barro et al.  2013,
2014a, 2014b; Marchesini et al. 2014; Schawinski et al.  2014; Belli et al.
2015; Wellons et al. 2015). 

\begin{acknowledgements}
We would like to thank the anonymous referee for helpful comments that improved the
manuscript. This project is supported by the NSF of China
11373027, 10973011 and 11003015.  The Project-sponsored by SRF for ROCS, SEM.
Funding for the creation and distribution of the SDSS Archive has been provided
by the Alfred P. Sloan Foundation, the Participating Institutions, the National
Aeronautics and Space Administration, the National Science Foundation, the U.S.
Department of Energy, the Japanese Monbukagakusho, and the Max Planck Society.
The SDSS Web site is http://www.sdss.org/.  The SDSS is managed by the
Astrophysical Research Consortium (ARC) for the Participating Institutions. The
Participating Institutions are The University of Chicago, Fermilab, the
Institute for Advanced Study, the Japan Participation Group, The Johns Hopkins
University, the Korean Scientist Group, Los Alamos National Laboratory, the
Max-Planck-Institute for Astronomy (MPIA), the Max-Planck-Institute for
Astrophysics (MPA), New Mexico State University, University of Pittsburgh,
Princeton University, the United States Naval Observatory, and the University
of Washington.  Some of the data presented in this paper were obtained from the
Mikulski Archive for Space Telescopes (MAST). STScI is operated by the
Association of Universities for Research in Astronomy, Inc., under NASA
contract NAS5-26555. Support for MAST for non-HST data is provided by the NASA
Office of Space Science via grant NNX09AF08G and by other grants and contracts.
\end{acknowledgements}


\clearpage

\label{lastpage}

\end{document}